\documentclass[final,5p,times,twocolumn]{elsarticle}

\usepackage{graphicx}
\usepackage{lineno}
\usepackage{url}
\usepackage{hyperref}

\journal{Physics Letters B}

\begin{document}

\biboptions{sort&compress} 

\begin{frontmatter}



\title{\vspace{-0.5cm} \flushright{\normalsize{CERN-PH-EP-2011-010}} \\ \center{Search for Massive Long-lived Highly Ionising Particles \\ with the ATLAS Detector at the LHC}}


\author{The ATLAS Collaboration\fnref{fn1}}

\fntext[fn1]{See Appendix for the list of collaboration members}

\begin{abstract}
A search is made for massive highly ionising particles with lifetimes in excess of 100 ns, with the ATLAS experiment at the Large Hadron Collider, using 3.1~pb$^{-1}$ of $pp$ collision data taken at $\sqrt{s}=7$~TeV. The signature of energy loss in the ATLAS inner detector and electromagnetic calorimeter is used. No such particles are found and limits on the production cross section for electric charges $6e\leq |q| \leq 17e$ and masses 200~GeV $\leq m \leq$ 1000~GeV are set in the range $1-12$~pb for different hypotheses on the production mechanism.
\end{abstract}


\end{frontmatter}

%
%

\section{Introduction}\label{introduction} 

The observation of a massive long-lived highly ionising particle (HIP) possessing a large electric charge $|q|\gg e$, where $e$ is the elementary charge, would represent striking evidence for physics beyond the Standard Model. Examples of putative particles which can give rise to HIP signatures include $Q$-balls~\cite{Kusenko:1997si}, stable micro black-hole remnants~\cite{Koch:2007}, magnetic monopoles~\cite{Fairbairn:2006gg} and dyons~\cite{Schwinger:1966nj}. Searches for HIPs are made in cosmic rays~\cite{Cecchini:2008su} and at colliders~\cite{Fairbairn:2006gg}; recent collider searches were performed at LEP~\cite{Kinoshita:1992wd,Pinfold:1993mq,OPALdirect} and the Tevatron~\cite{Price:1987py,Price:1990in,Bertani:1990tq,CDFdirect}. Cross sections and event topologies associated with HIP production cannot be reliably predicted due to the fact that the coupling between a HIP and the photon is so strong that perturbative calculations are not possible. Therefore, search results at colliders are usually quoted as cross section limits in a range of charge and mass for given kinematics~\cite{Fairbairn:2006gg}. Also, for the same reason, limits obtained at different collision energies or for different types of collisions cannot be directly compared; rather, they are complementary. 

HIP searches are part of a program of searches at the CERN Large Hadron Collider (LHC) which explore the multi-TeV energy regime. Further motivation is provided by the gauge hierarchy problem, to which proposed solutions typically postulate the existence of hitherto unobserved particles with TeV-scale masses. HIPs at the LHC can be sought at the dedicated MoEDAL plastic-track experiment~\cite{moedal} or, as in this work, via their active detection at a multipurpose detector. 

Due to their assumed large mass (hundreds of GeV), HIPs are characterised by their non-relativistic speed. The expected large amounts of energy loss per unit length (d$E/$d$x$) through ionisation (no bremsstrahlung) are mainly due to the high particle charge, but also due to the low speed. The ATLAS detector is well suited to detect HIPs. A HIP with sufficient kinetic energy would leave a track in the inner detector tracking system of ATLAS and lose its energy on its way to and through the electromagnetic calorimeter, giving rise to an electron-like signature. The presence of a HIP can be inferred from measurements of the proportion of high-ionisation hits in the inner detector. In addition, assuming isolation, the lateral extent of the energy deposition in the calorimeter is a sensitive discriminant between HIPs and Standard Model particles. 

The ranges of HIP charge, mass and lifetime for which unambiguous conclusions can be drawn are determined by the chosen trigger and event selections. The choice of an electromagnetic trigger limits the phase space to HIPs which stop in the electromagnetic calorimeter of ATLAS. The search is optimised for data collected at relatively low instantaneous luminosities (up to $10^{31}$~cm$^{-2}$s$^{-1}$), for which a low (10~GeV) trigger transverse energy threshold is available. In the barrel region of the calorimeter, this gives access to energy depositions corresponding to HIPs with electric charges down to $6e$. Standard electron reconstruction algorithms are used, which implies that tracks which bend like electrically charged particles are sought. Particles with magnetic charge, or electric charge above $17e$, are not addressed here due to the bending along the beam axis in the case of a monopole, and due to effects from delta electrons and electron recombination in the active detector at the corresponding values of energy loss (d$E/$d$x>2\cdot 10^3$~MeV/cm). For such types of HIPs, more detailed studies are needed to assess and minimise the impact of these effects on the selection efficiency. The 1000~GeV upper bound in mass sensitivity is determined by trigger timing constraints, as a significantly heavier HIP (with charge $17e$ or lower) would be delayed by more than 12~ns with respect to $\beta=1$ when it stops in the electromagnetic calorimeter (this corresponds to $\beta<0.3$), and would thus risk being triggered in the next proton bunch crossing. The search is sensitive to HIP lifetimes larger than $100$~ns since a particle which decays much earlier in the calorimeter (even after stopping) would spoil the signature of a narrow energy deposition. 

\section{The ATLAS Detector}\label{detector}

The ATLAS detector~\cite{Aad:2008zzm} is a multipurpose particle
physics apparatus with a forward-backward symmetric cylindrical
geometry and near 4$\pi$ coverage in solid angle~\cite{ATLASreferencesystem}.
A thin superconducting solenoid magnet surrounding the inner part of the ATLAS detector produces a field of approximately 2~T along the beam axis.

Inner detector (ID) tracking is performed by silicon-based detectors and an outer tracker using straw tubes with particle identification capabilities based on transition radiation (Transition Radiation Tracker, TRT). The TRT is divided into barrel (covering the pseudorapidity range $|\eta|<1.0$) and endcap ($0.8<|\eta|<2.0$) components. A track gives a typical number of straw hits of 36. At the front-end electronics of the TRT, discriminators are used to compare the signal against low and high thresholds. While the TRT has two hit threshold levels, there is no upper limit to the amount of ionisation in a straw which will lead to a signal~\cite{ATLASTRT}, guaranteeing that highly ionising particles would not escape detection in the TRT. Rather, they would produce a large number of high-threshold (HT) hits along their trajectories. The amount of ionisation in a straw tube  needed for a TRT HT hit is roughly equivalent to three times that expected from a minimum ionising particle.  

Liquid-argon sampling electromagnetic (EM) calorimeters, which comprise accordion-shaped electrodes and lead absorbers, surround the ID. The EM calorimeter barrel ($|\eta|<1.475$) is used in this search. It is segmented transversely and divided in three layers in depth, denoted first, second, and third layer, respectively. In front of the accordion calorimeter a thin presampler layer is used to correct for fluctuations of energy loss. The typical cell granularity ($\Delta\eta\times\Delta\phi$) of the EM barrel is $0.003\times 0.1$ in the first layer and $0.025\times 0.025$ in the second layer. The signal expected for a HIP in the considered charge range lies in a region in time and energy where the electronic response in EM calorimeter cells is well understood and does not saturate. The robustness of the EM calorimeter energy reconstruction has been studied in detail and pulse shape predictions are consistent with the measured signals~\cite{ATLASLAr}. 

The stopping power of a HIP in the ATLAS detector depends on its charge, mass and energy, as well as the material budget along its path. Details of the latter are given in Ref. \cite{:2008zzm} in terms of number of radiation lengths $X_0$, as a function of depth and pseudorapidity. The integrated radiation length between the interaction point and the exit of the TRT is 0.5 $X_0$ at $\eta=0$ and 1.5 $X_0$ at $|\eta|=1.3$. The additional amount of material before the first layer of the EM calorimeter is 2.0 $X_0$ at $\eta=0$ and 3.5 $X_0$ at $|\eta|=1.3$. The thicknesses of the first, second and third EM layers are 4.5 $X_0$, 16.5 $X_0$ and 1.5 $X_0$ at $\eta=0$ and 3 $X_0$, 20 $X_0$ and 5 $X_0$ at $|\eta|=1.3$, respectively.

\section{Simulated Event Samples}\label{simulation}

\begin{figure}[Ht]
  \begin{center}
    \includegraphics[width=0.98\linewidth]{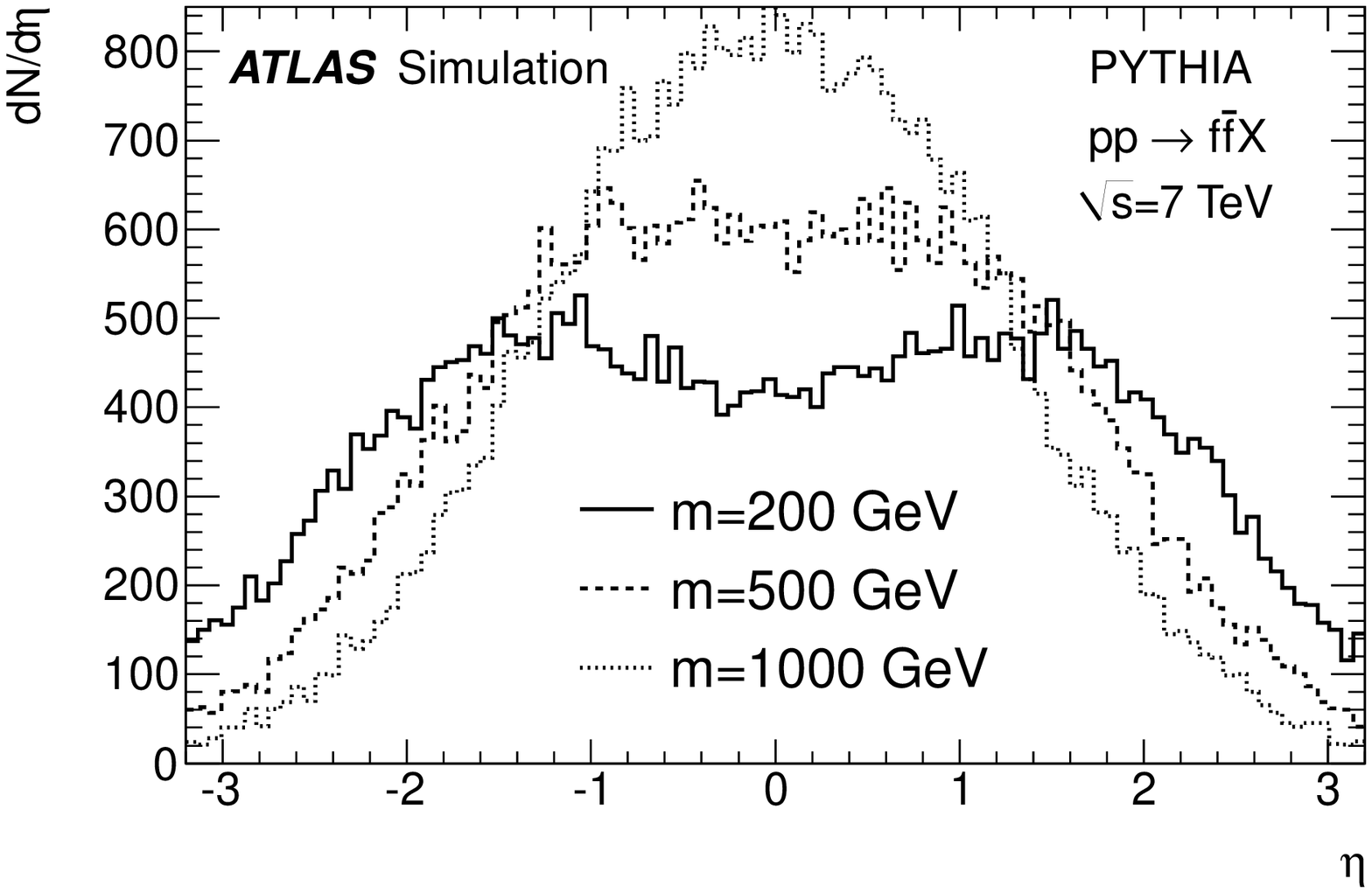}
    \includegraphics[width=0.98\linewidth]{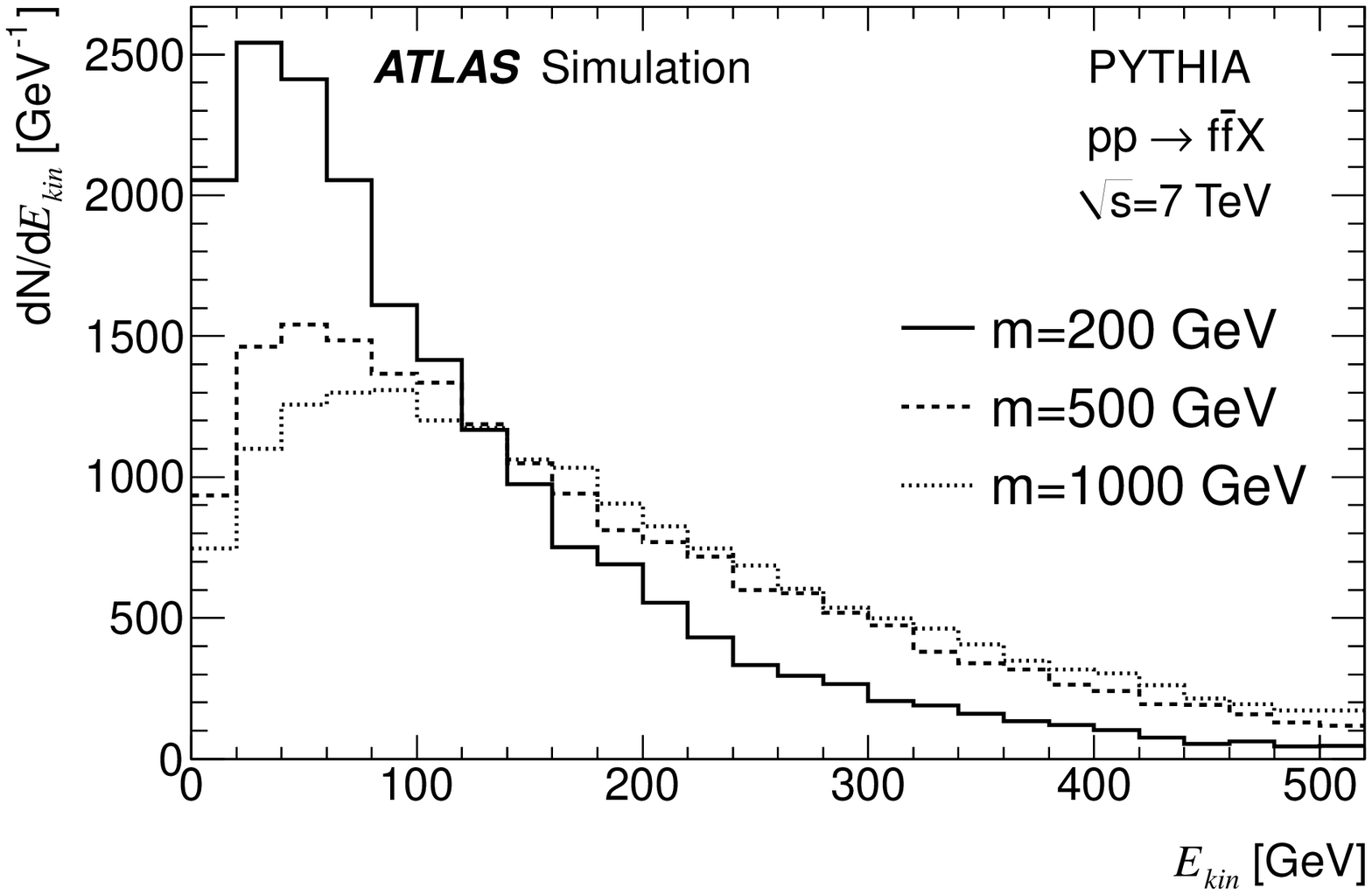}  
  \end{center}
  \caption{Distributions of pseudorapidity $\eta$ (top) and kinetic energy $E_{kin}$ (bottom) at origin for heavy fermions produced with the Drell-Yan process. The latter is given with a requirement of $|\eta|<1.35$. The distributions for the three different masses are normalised to the same number of entries.}
  \label{fig:Ekin}
\end{figure}

Signal events are generated with the {\sc Pythia} Monte Carlo (MC) event generator~\cite{Sjostrand:2006za} according to the fermion pair production process: $p+p\rightarrow f+\bar{f}+X$. Ref.~\cite{Sherstnev:2007nd} is used for the parton distributions of the proton. Direct pair production implies that the HIPs are not part of a jet and thus isolated. A Drell-Yan-like production mechanism, modified to take into account the mass of the HIP~\cite{Kalbfleisch:2003yt}, is used to model the kinematic properties of the HIPs. Generated $\eta$ distributions, as well as kinetic energy ($E_{kin}$) spectra in the central region ($|\eta|<1.35$), are shown in Figure~\ref{fig:Ekin} for the three mass points considered in this search. 

An ATLAS detector simulation~\cite{ATLASG4} based on {\sc Geant-4}~\cite{Agostinelli:2002hh} is used, where the particle interactions include secondary ionisation by delta electrons in addition to the standard ionisation process based on the Bethe-Bloch formula. A correction for electron-ion recombination effects in the EM calorimeter (Birks' correction) is applied, with typical visible energy fractions between 0.2 and 0.5 for the signal particles considered. Effects of delays are simulated, except for the ability to trigger slow-moving particles within the proton bunch crossing time, which is considered separately as a systematic uncertainty (see Section \ref{sec:systematics}). Samples of approximately 20000 events are produced for HIPs with masses of 200, 500 and 1000~GeV. For each mass point, HIPs with charges $6e$, $10e$ and $17e$ are simulated. 

A data-driven method is used in this work to estimate backgrounds surviving the final selections (see Section~\ref{sec:bg}). However, in order to demonstrate that the distributions of the relevant observables are understood, a sample of simulated background events is used. The background sample, generated with {\sc Pythia}~\cite{Sjostrand:2006za} and labeled ``Standard Model'', consists mostly of QCD events in which the hard subprocess is a strong 2-to-2 process with a matrix element transverse momentum cut-off of 15~GeV, but also includes contributions from heavy quark and vector boson production. A true transverse energy larger than 17~GeV in a typical first level trigger tower is also required. This sample contains $4\cdot 10^7$ events and corresponds roughly to an integrated luminosity of $0.8$~pb$^{-1}$.

\section{Trigger and Event Selection}\label{selection}

The collected data sample corresponds to an integrated luminosity of $3.1\pm0.3$~pb$^{-1}$, using a first level trigger based on energy deposits in the calorimeters. At the first level of the trigger, so-called trigger towers with dimension $\Delta\eta\times\Delta\phi=0.1 \times 0.1$ are defined. In each trigger tower the cells of the electromagnetic or hadronic calorimeter are summed. EM clusters with fixed size $\Delta\eta\times\Delta\phi=0.2 \times 0.2$ are sought and are retained if the total transverse energy ($E_T$) in an adjacent pair of their four trigger towers is above 5~GeV. Further electron-like higher level trigger requirements are imposed on the candidate, including $E_T>10$~GeV, a matching to a track in the ID and a veto on hadronic leakage~\cite{atlastrigger}. The efficiency of this trigger for the data under consideration is measured to be $(94.0\pm1.5)\%$ for electrons with $E_T>15$~GeV and is well described by the simulation. The simulation predicts that a highly charged particle which stops in the EM barrel would be triggered with a similar efficiency or higher.

Offline electron candidates have cluster sizes of $\Delta\eta\times\Delta\phi=0.075\times 0.175$ in the EM barrel, with a matched track in a window of $\Delta\eta\times\Delta\phi=0.05\times 0.1$ amongst reconstructed tracks with transverse momentum larger than $0.5$~GeV. 
Identification requirements corresponding to ``medium'' electrons~\cite{WZATLAS2010}, implying track and shower shape quality cuts, are applied to the candidates. These cuts filter out backgrounds but have a negligible impact on the signal, for which the cluster width is much narrower than for typical electrons. The cluster energy is estimated correcting for the energy deposited outside the active calorimeter regions, assuming an EM shower.

Further offline selections on the cluster transverse energy ($E_T>15$~GeV) and pseudorapidity ($|\eta|<1.35$) are imposed. The $E_T$ selection guarantees that the trigger efficiency is higher than $94\%$ for the objects under study. 
The restriction of $|\eta|<1.35$ excludes the transition region between the EM calorimeter barrel and endcap, reducing the probability for backgrounds to fake a narrow energy deposition. 

\subsection{Selection Cuts}\label{sec:sel}

A loose selection based on TRT and EM calorimeter information is also imposed on the candidates to ensure that the quality of the track and cluster associated to the electron-like object is good enough to ensure the robustness of the HIP selection variables, and to provide a data sample with which to estimate the background rate. Only candidates with more than 10 TRT hits are retained. In addition to the $E_T>15$~GeV cut for the EM cluster associated with the candidate, a significant fraction of the total cluster energy is required to be contained in six calorimeter cells among the first and second EM layers. This is done by requiring the summed energy in the three most energetic cells in each of the first and second layers to be greater than 2 and 4~GeV, respectively. Following these selections, $137503$ candidates remain in the data.

Two sets of observables are used in the final selection. The ID-based observable is the fraction, $f_{HT}$, of TRT hits on the track which pass the high threshold. The calorimeter-based discriminants are the fractions of energies outside of the three most energetic cells associated to a selected EM cluster, in the first and second EM calorimeter layers: $w_1$ and $w_2$.

\begin{figure}[ht]
  \begin{center}
    \includegraphics[width=0.98\linewidth]{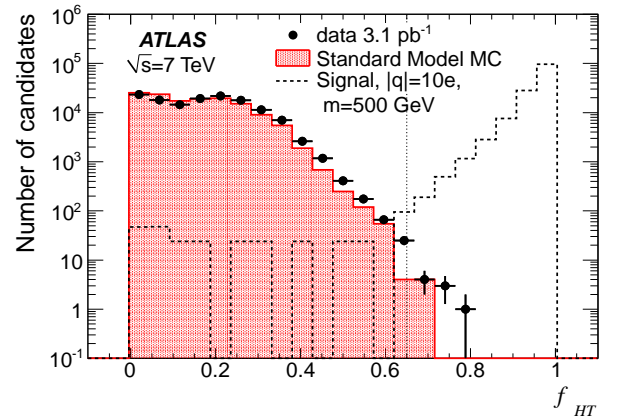}
  \end{center}
  \caption{Distribution of the fraction of TRT high-threshold hits for candidates satisfying the loose selection. Data (dots) are compared with area-normalised signal ($|q|=10e$ and $m=500$~GeV, dashed line) and Standard Model background (shaded area) simulations. The dotted line shows the selection cut value.}
  \label{fig:hthits}
\end{figure}

The $f_{HT}$ distribution for loosely selected candidates is shown in Figure~\ref{fig:hthits}. The data extend up to $f_{HT}=0.8$. The prediction of the signal simulation for a HIP of mass 500~GeV and charge $10e$ is also shown. It peaks at $f_{HT}\sim1$ and has a small tail extending into the Standard Model region.

\begin{figure}[Ht]
  \begin{center}
    \includegraphics[width=0.98\linewidth]{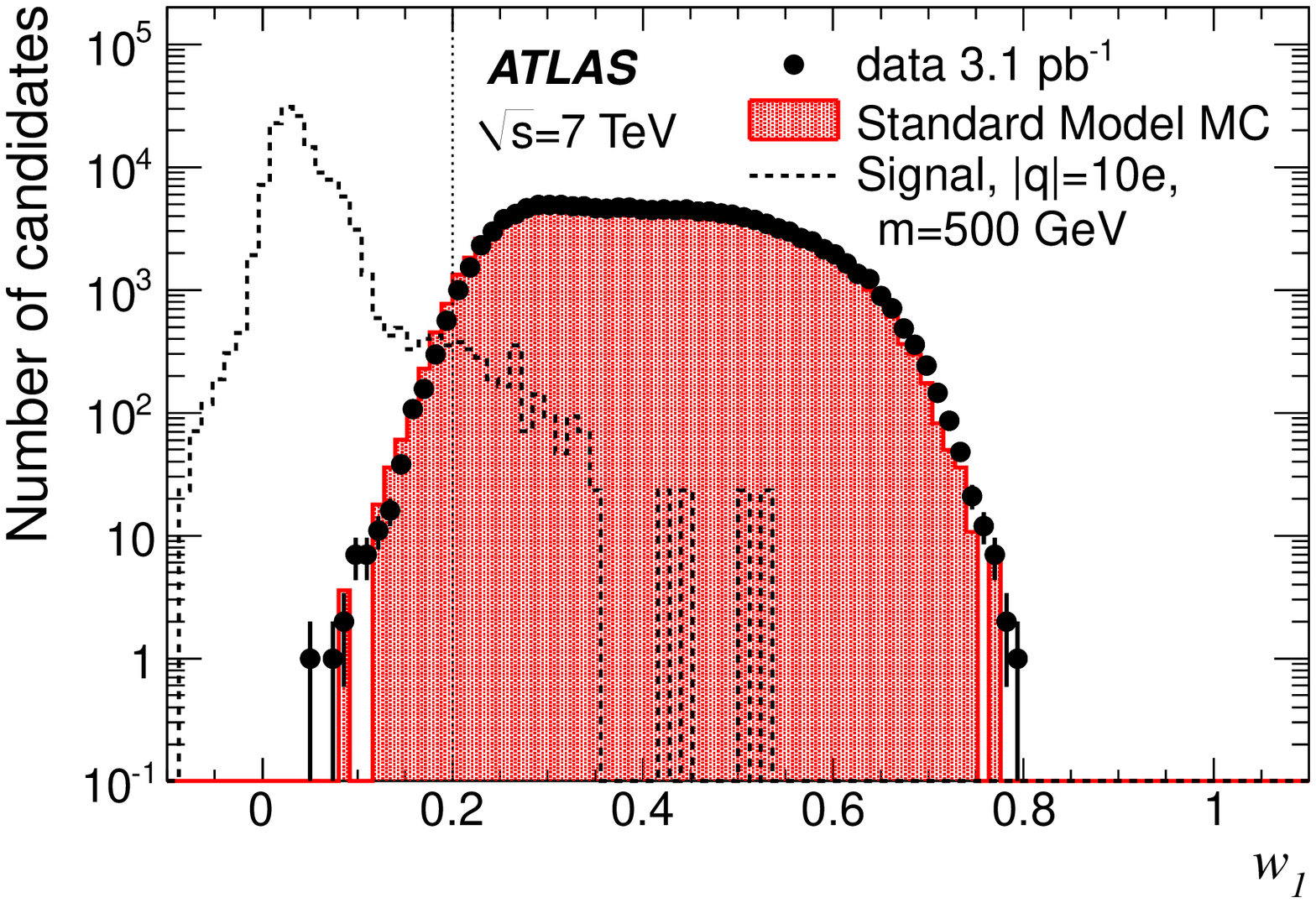}
    \includegraphics[width=0.98\linewidth]{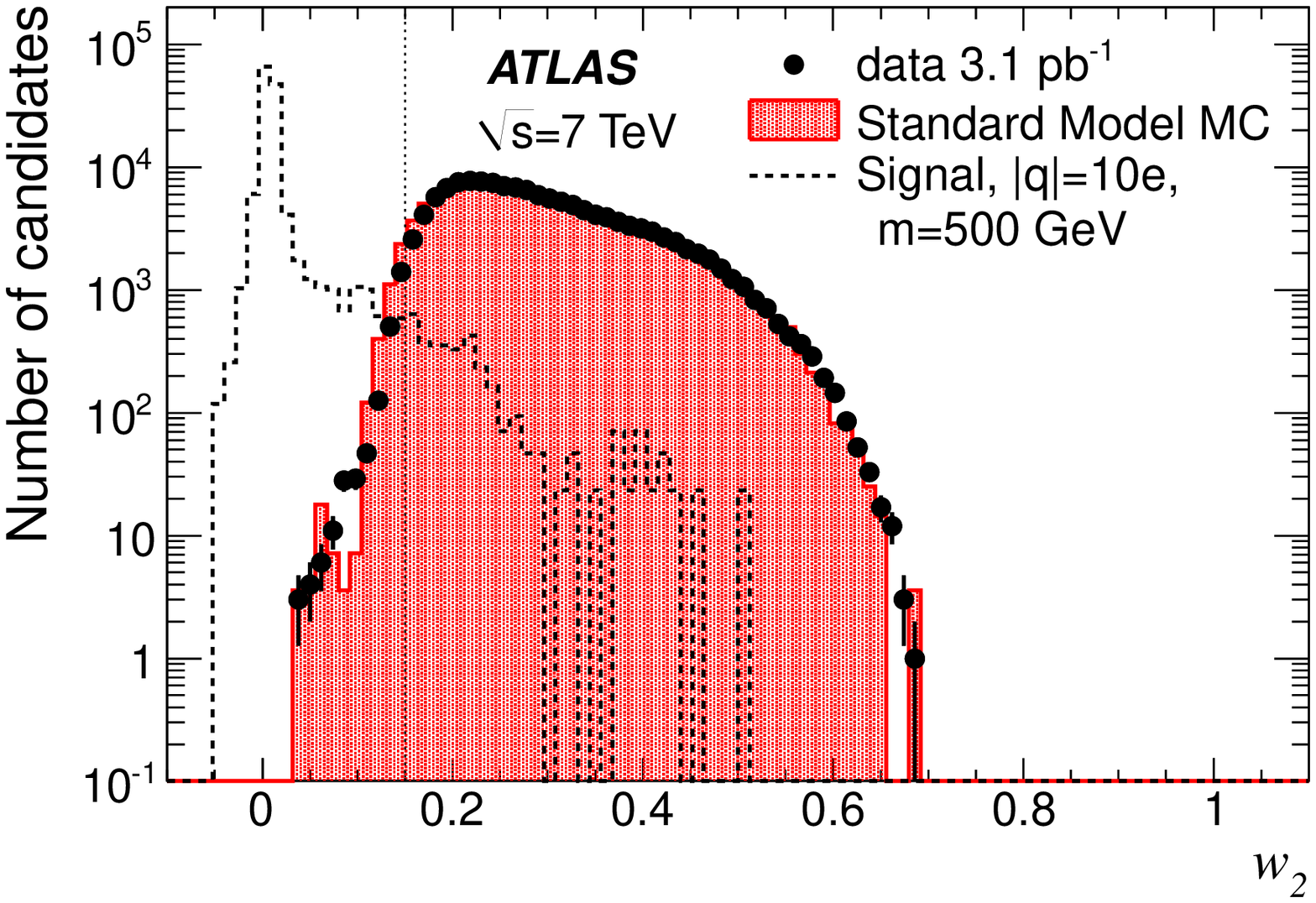}
  \end{center}
  \caption{Distributions of $w_1$ and $w_2$ following the loose selection. Data (dots) are compared with area-normalised  signal ($|q|=10e$ and $m=500$~GeV, dashed lines) and Standard Model background (shaded area) simulations. Negative values are caused by pedestal fluctuations. Dotted lines show the selection cut values.}
  \label{fig:figw}
\end{figure}

The distributions of $w_1$ and $w_2$ also provide good discrimination between signal and background, as shown in Figure~\ref{fig:figw}. For a signal, energy is deposited only in the few cells along the particle trajectory (as opposed to backgrounds which induce showers in the EM calorimeter) and the distributions peak around zero for both variables. The shapes of the measured distributions are well described by the background simulation. A faint double-peak structure is visible in data and in background simulations for the $f_{HT}$, $w_1$ and $w_2$ distributions in Figures \ref{fig:hthits} and \ref{fig:figw}, where the main peak (closest to the signal) corresponds to electrons and the secondary peak corresponds to hadrons which fake the electron identification signature.

Finally, the following HIP selection is made: $f_{HT}>0.65$, $w_1<0.20$ and $w_2<0.15$. For signal particles, these cuts reject only candidates in the tails of the distributions, and varying them has a minor impact on the efficiency; this feature is common to all considered charge and mass points. The cut values were chosen to yield a very small ($\ll 1$ event) expected background (see Section~\ref{sec:bg}) while retaining a high ($\sim9
6\%$) efficiency for the signal. No candidates in data or in simulated Standard Model events pass this selection.  

\subsection{Data-driven Background Estimation}\label{sec:bg}

\begin{figure}[ht]
  \begin{center}
    \includegraphics[width=0.98\linewidth]{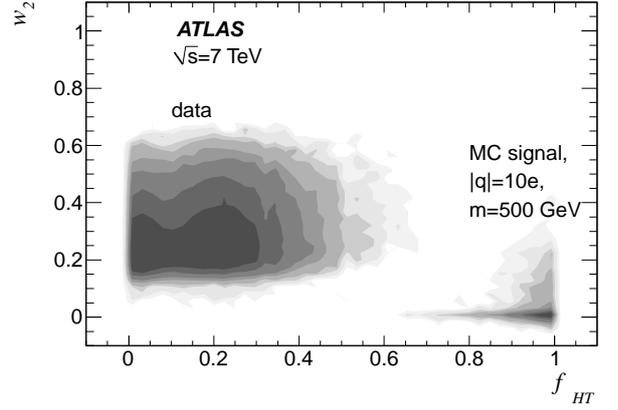}
  \end{center}
  \caption{Contours of $w_2$ versus $f_{HT}$ distributions following loose selection, showing the density of entries on a log scale. Data and signal Monte Carlo ($|q|=10e$ and $m=500$~GeV) are shown, and no candidates in the data appear near the signal region. The correlation factor between $w_2$ and $f_{HT}$ in the data is positive (coefficient 0.15); the same trend is also true for the correlation between $w_1$ and $f_{HT}$ (coefficient 0.18).}
  \label{fig:w_fHT}
\end{figure}

A data-driven method is used to quantify the expected background yield after the HIP selection. Potential backgrounds consist mainly of electrons. For Standard Model candidates, the ID and calorimeter observables are correlated in a way that further suppresses the backgrounds (see Figure~\ref{fig:w_fHT}). The background estimation assumes that $f_{HT}$ is uncorrelated with $w_1$ and $w_2$ and is thus conservative. 

The yield of particle candidates passing the loose selection $N_{loose}=137503$ can be divided into the following: $N_0$, $N_1$, $N_{f_{HT}}$, and $N_w$, which represent the number of candidates which satisfy both of the selections, neither of the selections, only the $f_{HT}$ selection, and only the $w_1$ and $w_2$ selections taken together, respectively. Even in the presence of a signal, $N_1$, $N_{f_{HT}}$ and $N_w$ would be dominantly composed of background events. The probability of a background candidate passing the TRT requirement is then $P_{f_{HT}}=\frac{N_{f_{HT}}}{(N_1+N_{f_{HT}})}$ and the probability to pass the calorimeter requirements is $P_{w}=\frac{N_w}{(N_1+N_w)}$, leading to an expectation of the number of background candidates entering the signal region: $N_{bg}=N_{loose}P_{f_{HT}}P_{w}$. The data sample yields $N_0=0$, $N_1=137342$, $N_{f_{HT}}=18$ and $N_w=143$, leading to $P_{f_{HT}}=(1.3\pm0.3)\cdot 10^{-4}$ and $P_{w}=(1.0\pm0.1)\cdot 10^{-3}$. The expected number of background candidates surviving the selection, and thereby the expected number of background events, is thus $N_{bg}=0.019\pm0.005$. The quoted uncertainty is statistical. 

\section{Signal Selection Efficiency}\label{efficiency} 

\subsection{Efficiencies in Acceptance Kinematic Regions}

\begin{figure}[ht]
  \begin{center}
    \includegraphics[width=0.98\linewidth]{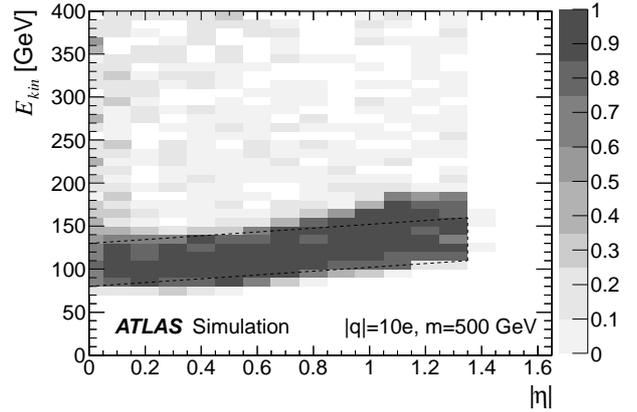}
  \end{center}
  \caption{Probability to pass all selection criteria as a function of pseudorapidity and kinetic energy at origin, for a HIP with charge $10e$ and mass 500~GeV. The dark region corresponds to the kinetic range where the particle stops in or near the second layer of the EM calorimeter barrel and is parametrised with three energy values (dashed parallelogram, see Table~\ref{tab:Eranges}).}
  \label{fig:effmap}
\end{figure}

The probability to retain a signal event can be factorised in two parts: acceptance (probability for a HIP in a region where the detector is sensitive) and efficiency (probability for this HIP to pass the selection cuts). The acceptance is defined here as the probability that at least one signal particle will be in the range $|\eta|<1.35$ and stop in the second or third layer of the EM calorimeter. If this condition is satisfied, the simulation predicts a high probability to trigger on, reconstruct and select the event. This corresponds to the dark region in Figure~\ref{fig:effmap}, which shows the predicted selection efficiency mapped as a function of the initial HIP pseudorapidity and kinetic energy, in the case of $|q|=10e$ and $m=500$~GeV. Such acceptance kinematic regions can be parametrised with three values defining three corners of a parallelogram. These parameters are summarised in Table~\ref{tab:Eranges}. For HIPs produced inside such regions, the candidate selection efficiency is flat within 10\% and takes values between $0.5$ and $0.9$ depending on the charge and mass (see Table~\ref{tab:effmap}). For $|q|=17e$, the main source of inefficiency is the requirement on the number of TRT HT hits, which contributes up to 20\% signal loss. This is largely due to the presence of track segments from delta electrons, which have a non-negligible probability to be chosen by the standard electron track matching algorithm. For low charges, inefficiencies are dominated by the cluster $E_T$ cut, typically accounting for $\sim6$\% loss. Other contributions, like trigger, electron reconstruction, and electron identification, can each cause $1-6\%$ additional inefficiency. 

\begin{table}
  \begin{center}
    \begin{tabular}{cc|ccc}
                               $|q|$ & $m$ [GeV]           & $E_{kin}^{min}$     & $E_{kin}^{min}$        &  $E_{kin}^{max}$ \\
                                    && $(\eta=0)$          & $(|\eta|=1.35)$        &  $(\eta=0)$\\
      \hline
      $6e$ & $200$          & 40                      & 50                      & 50   \\
      $6e$ & $500$          & 50                      & 70                      & 70   \\
      $6e$ & $1000$         & 60                      & 130                     & 80   \\
      $10e$ & $200$         & 50                      & 80                      & 90   \\
      $10e$ & $500$         & 80                      & 110                     & 130  \\
      $10e$ & $1000$        & 110                     & 150                     & 180  \\
      $17e$ & $200$         & 100                     & 150                     & 190  \\
      $17e$ & $500$         & 150                     & 190                     & 260  \\
      $17e$ & $1000$        & 190                     & 240                     & 350  \\
    \end{tabular}
  \caption{Kinetic energies (in GeV) defining the acceptance kinematic ranges for HIPs with the masses and electric charges considered in this search. The three columns correspond to the lower left, lower right, and upper left corners of parallelograms in the $\left(|\eta|, E_{kin}\right)$ plane.}
  \label{tab:Eranges}
  \end{center}
\end{table}

\begin{table}
  \begin{center}
    \begin{tabular}{l|ccc}
            $m$ [GeV]             & $|q|=6e$               & $|q|=10e$              & $|q|=17e$  \\ 
      \hline
            $200$          & $0.822\pm 0.026$      & $0.820\pm 0.015$      & $0.484\pm 0.012$ \\
            $500$          & $0.868\pm 0.021$      & $0.856\pm 0.014$      & $0.617\pm 0.011$ \\
            $1000$         & $0.558\pm 0.019$      & $0.858\pm 0.012$      & $0.700\pm 0.012$ \\
    \end{tabular}
  \caption{Expected fractions of HIP candidates passing the final selection, assuming they are isolated and produced inside the acceptance regions defined by the values in Table~\ref{tab:Eranges}. Uncertainties due to MC statistics are quoted; other systematic uncertainties are discussed in Section~\ref{sec:systematics}.}
  \label{tab:effmap}
  \end{center}
\end{table}

\subsection{Efficiencies for Drell-Yan Kinematics}

The estimated fractions of signal events where at least one candidate passes the final selection, assuming they are produced with Drell-Yan kinematics, are shown in Table~\ref{tab:effDY} for the values of charge and mass considered in this search. 
The dominant source of loss ($70-85$\% loss) is from the kinematic acceptance, i.e., the production of HIPs with $|\eta|>1.35$, as well as their stopping before they reach the second layer of the EM calorimeter, or after they reach the first layer of the hadronic calorimeter. The relative contributions from these various types of acceptance loss depend on mass and charge, as well as the kinematics of the assumed production model. The Drell-Yan production model implies that the fraction of HIPs produced in the acceptance region of pseudorapidity $|\eta|<1.35$ is larger with increasing mass (see Figure~\ref{fig:Ekin}). Also, with the assumed energy spectra (bottom plot in Figure~\ref{fig:Ekin}), the acceptance is highest for intermediate charges ($|q|=10e$), since HIPs with low charges tend to punch through the EM calorimeter and HIPs with high charges tend to stop before reaching it. 

\begin{table}
  \begin{center}
    \begin{tabular}{l|ccc}
       $m$ [GeV]        & $|q|=6e$               & $|q|=10e$              & $|q|=17e$  \\ 
      \hline
      $200$         & $0.102\pm 0.002$    & $0.175\pm 0.003$    & $0.112\pm 0.002$ \\
      $500$         & $0.150\pm 0.003$    & $0.236\pm 0.003$    & $0.193\pm 0.003$ \\
      $1000$        & $0.133\pm 0.002$    & $0.299\pm 0.004$    & $0.237\pm 0.004$ \\
    \end{tabular}
  \caption{Expected fractions of signal events passing the final selection, assuming Drell-Yan kinematics. Uncertainties due to MC statistics are quoted; other systematic uncertainties are discussed in Section~\ref{sec:systematics}.}
  \label{tab:effDY}
  \end{center}
\end{table}

\section{Systematic Uncertainties}\label{sec:systematics}

The major sources of systematic uncertainties affecting the efficiency estimation are summarised below. These mainly concern possible imperfections in the description of HIPs in the detector by the simulation. 

\begin{itemize}
\item  The recombination of electrons and ions in the sampling region of the EM calorimeter affects the measured current and thus the total visible energy. Recombination effects become larger with increasing  d$E$/d$x$. In the ATLAS simulation, this is parametrised by Birks' law~\cite{Birkslaw}. To estimate the uncertainty associated with the approximate modeling of recombination effects, predictions from the ATLAS implementation of Birks' correction~\cite{LArResponse} are compared to existing data of heavy ions punching through a layer of liquid argon~\cite{LArHIDoke,LArHIShibamura,LArHICrawford}. In the range $2\cdot 10^2$~MeV/cm $<$ d$E$/d$x$ $<2\cdot 10^3$~MeV/cm, which corresponds to typical HIP energy losses in the EM calorimeter for the charges and masses under consideration, the uncertainty in the simulated visible energy fraction is $\pm15\%$. This introduces between $4$\% and $23$\% uncertainty in the signal selection efficiency. The impact is largest for charge $6e$, for which a lower visible energy would be more likely to push the candidate below the 15~GeV cluster $E_T$ threshold. 
\item The fraction of HIPs which stop in the detector prior to reaching the EM calorimeter is affected by the assumed amount of material in the {\sc Geant-4} simulation. Varying the material density within the assumed uncertainty range ($\pm\sim10\%$~\cite{ATLASChargedMulti}), independently in the ID and EM calorimeter volumes, leads to a 6\% uncertainty in signal acceptance.
\item The modeling of inactive or inefficient EM calorimeter regions in the simulation results in a 2\% uncertainty in the signal efficiency. 
\item Cross-talk effects between EM calorimeter cells affect the $w_1$ and $w_2$ variables and this may not be accurately described by the simulation for large energy depositions per cell. The resulting uncertainty in signal efficiency is 2\%. 
\item Secondary ionisation by delta electrons affects the track reconstruction and the calorimeter energy output. The amount of delta electrons in ATLAS detectors as described in {\sc Geant-4} depends on the cutoff parameter (the radius beyond which delta electrons are considered separate from the mother particle). Varying this parameter results in a 3\% uncertainty in the signal efficiency.
\item For clusters delayed by more than 10~ns with respect to the expected arrival time of a highly relativistic particle, which corresponds to $\beta<0.37$, there is a significant chance that the event is triggered in the next bunch crossing by the first level EM trigger. In most of the mass and charge range considered in this search, more than 99\% of the particles which are energetic enough to reach the EM calorimeter and pass the event selection are in the high-efficiency range $\beta>0.4$. The only exception is $|q|=6e$ and $m=1000$~GeV, for which the $\beta$ distribution after selection peaks between 0.32 and 0.47. The trigger efficiency loss is corrected for, resulting in an additional 25\% uncertainty for this particular case. 
\item Uncertainties in the choice of parametrisation for the parton density functions (pdfs) of the proton have an impact on the event kinematics. To test this effect, events were generated (see Section~\ref{simulation}) with 7 different pdfs from various sources~\cite{Sherstnev:2007nd,Gluck:1994uf,Lai:1999wy,Pumplin:2002vw,Martin:2009iq}. Assuming that acceptance variations due to the choice of pdf are Gaussian, the resulting relative uncertainty in the acceptance is 3\%. 
\item The relative uncertainty in efficiency due to MC statistics is of the order of 2\%.
\end{itemize}

Other effects, like event pile-up and electron pick-up by positively charged particles, have been investigated and found to be negligible. Efficiency systematics are dominated by Birks' correction. The relative uncertainties in the signal selection efficiencies (Tables~\ref{tab:effmap} and~\ref{tab:effDY}), obtained by adding all effects in quadrature, are shown in Table~\ref{tab:sysall}. 

The systematic uncertainty in the absolute integrated luminosity is 11\%~\cite{ATLASLumi}.  

\begin{table}
  \begin{center}
    \begin{tabular}{c|ccc}
      $m$ [GeV] & $|q|=6e$ & $|q|=10e$ & $|q|=17e$ \\
      \hline
      $200$    & 25\%  & 11\%  &  9\%   \\
      $500$    & 17\%  & 10\%  &  9\%   \\
      $1000$   & 28\%  & 10\%  &  9\%   \\
    \end{tabular} 
    \caption{Relative systematic uncertainties in efficiency, combining in quadrature all the effects described in the text.}
    \label{tab:sysall}
  \end{center}
\end{table}

\section{Upper Limit on the Cross Section}\label{limits}

A very low ($\ll 1$ event) background yield is expected and no events are observed to pass the selection. 
Knowing the integrated luminosity (3.1~pb$^{-1}$) and the selection efficiency for various model assumptions (Tables~\ref{tab:effmap} and~\ref{tab:effDY}), cross section limits are obtained. This is done using a Bayesian statistical approach with a uniform prior for the signal and the standard assumption that the uncertainties in integrated luminosity (11\%) and efficiency (Table~\ref{tab:sysall}) are Gaussian and independent. The limits are presented in Table~\ref{tab:xskin} (for a particle produced in the acceptance kinematic region defined by Table~\ref{tab:Eranges}) and in Table~\ref{tab:xsDY} (assuming Drell-Yan kinematics). 

These limits can be approximately interpolated to intermediate values of mass and charge. Also, the limits quoted in Table~\ref{tab:xskin} can be used to extract cross section limits for any given model of kinematics by correcting for the acceptance (fraction of events with at least one generated HIP in the ranges defined by Table~\ref{tab:Eranges}): such a procedure yields conservative limits thanks to the fact that candidates beyond the sharp edges of the acceptance regions defined in Table~\ref{tab:Eranges} can also be accepted.

\begin{table}
\begin{center}
\begin{tabular}{l|ccc}
$m$ [GeV] & $|q|=6e$ & $|q|=10e$ & $|q|=17e$ \\
\hline
$200$   & $1.4$ & $1.2$ & $2.1$ \\
$500$   & $1.2$ & $1.2$ & $1.6$ \\
$1000$  & $2.2$ & $1.2$ & $1.5$ \\
\end{tabular}
\caption{Inclusive HIP cross section upper limits (in pb) at 95\% confidence level for isolated long-lived massive particles with high electric charges produced in regions of pseudorapidity and kinetic energy as defined in Table~\ref{tab:Eranges}. Efficiencies in Table~\ref{tab:effmap} and uncertainties in Table~\ref{tab:sysall} were used in the cross section limit calculation. }
\label{tab:xskin}
\end{center}
\end{table}

\begin{table}
  \begin{center}
    \begin{tabular}{l|ccc}
$m$ [GeV]            & $|q|=6e$ & $|q|=10e$ & $|q|=17e$  \\
      \hline
$200$   & $11.5$ & $5.9$ & $9.1$ \\
$500$   & $7.2$  & $4.3$ & $5.3$ \\
$1000$  & $9.3$  & $3.4$ & $4.3$ \\
    \end{tabular}
  \caption{Pair production cross section upper limits (in pb) at 95\% confidence level for long-lived massive particles with high electric charges, assuming a Drell-Yan mechanism. Efficiencies in Table~\ref{tab:effDY} and uncertainties in Table~\ref{tab:sysall} were used in the cross section limit calculation. }
  \label{tab:xsDY}
  \end{center}
\end{table}

\section{Summary}\label{summary}

A search has been made for HIPs with lifetimes in excess of 100 ns produced in the ATLAS detector at the LHC using 3.1~pb$^{-1}$ of $pp$ collisions at $\sqrt{s} = 7$~TeV. The signature of high ionisation in an inner detector track matched to a narrow calorimeter cluster has been used. Upper cross section limits between 1.2~pb and 11.5~pb have been extracted for HIPs with electric charges between $6e$ and $17e$ and masses between 200~GeV and 1000~GeV, under two kinematics assumptions: a generic isolated HIP in a fiducial range of pseudorapidity and kinetic energy, or a Drell-Yan fermion pair production mechanism. HIP mass ranges above 800~GeV~\cite{Bertani:1990tq} are probed for the first time at a particle collider. These limits are the first constraints obtained on long-lived highly charged particle production at LHC collision energies.

%
%
%
%

\bibliographystyle{atlasstylem}
\bibliography{atlaspaper}

\begin{thebibliography}{10}
\bibitem{ATLASreferencesystem}
 ATLAS uses a right-handed coordinate system with its origin at the nominal
  interaction point (IP) in the centre of the detector and the $z$-axis
  coinciding with the axis of the beam pipe. The $x$-axis points from the IP to
  the centre of the LHC ring, and the $y$ axis points upward. Cylindrical
  coordinates $(r,\phi)$ are used in the transverse plane, $\phi$ being the
  azimuthal angle around the beam pipe. The pseudorapidity is defined in terms
  of the polar angle $\theta$ as $\eta=-\ln\tan(\theta/2)$.
\end{thebibliography}



\newpage

\begin{flushleft}
{\Large The ATLAS Collaboration}

\bigskip

G.~Aad$^{\rm 48}$,
B.~Abbott$^{\rm 111}$,
J.~Abdallah$^{\rm 11}$,
A.A.~Abdelalim$^{\rm 49}$,
A.~Abdesselam$^{\rm 118}$,
O.~Abdinov$^{\rm 10}$,
B.~Abi$^{\rm 112}$,
M.~Abolins$^{\rm 88}$,
H.~Abramowicz$^{\rm 153}$,
H.~Abreu$^{\rm 115}$
% ...
\bigskip

% ...
$^{10}$ Institute of Physics, Azerbaijan Academy of Sciences, H. Javid Avenue 33, AZ 143 Baku, Azerbaijan\\
$^{11}$ Institut de F\'isica d'Altes Energies, IFAE, Edifici Cn, Universitat Aut\`onoma  de Barcelona,  ES - 08193 Bellaterra (Barcelona), Spain\\
% ...
$^{48}$ Albert-Ludwigs-Universit\"{a}t, Fakult\"{a}t f\"{u}r Mathematik und Physik, Hermann-Herder Str. 3, D - 79104 Freiburg i.Br., Germany\\
$^{49}$ Universit\'e de Gen\`eve, Section de Physique, 24 rue Ernest Ansermet, CH - 1211 Geneve 4, Switzerland\\
% ...
$^{88}$ Michigan State University, Department of Physics and Astronomy, High Energy Physics Group, East Lansing, MI 48824-2320, United States of America\\
% ...
$^{111}$ University of Oklahoma, Homer L. Dodge Department of Physics and Astronomy, 440 West Brooks, Room 100, Norman, OK 73019-0225, United States of America\\
$^{112}$ Oklahoma State University, Department of Physics, 145 Physical Sciences Building, Stillwater, OK 74078-3072, United States of America\\
% ...
$^{115}$ LAL, Univ. Paris-Sud, IN2P3/CNRS, Orsay, France\\
% ...
$^{118}$ Oxford University, Department of Physics, Denys Wilkinson Building, Keble Road, Oxford OX1 3RH, United Kingdom\\
% ...
$^{153}$ Tel Aviv University, Raymond and Beverly Sackler School of Physics and Astronomy, Ramat Aviv, IL - Tel Aviv 69978, Israel\\
% ...
\end{flushleft}

\end{document}